\documentclass[iop]{emulateapj}

\usepackage{graphicx}  % needed for figures
\graphicspath{{./figures/}}
\usepackage{bm}        % for math
\usepackage{amsfonts,amsmath,amssymb,mathrsfs}
\usepackage{color}
\usepackage{ulem}

\usepackage{hyperref}
\hypersetup{
  colorlinks=true,        % false: boxed links; true: colored links
  linkcolor=blue,         % color of internal links
  citecolor=cyan,         % color of links to bibliography
}
\usepackage{natbib,times}
\definecolor{orange}{rgb}{1,0.5,0}

\newcommand{\ie}{\textit{i.e.,}}
\newcommand{\eg}{\textit{e.g.,}}

\def\gap{\;\rlap{\lower 2.5pt
\hbox{$\sim$}}\raise 1.5pt\hbox{$>$}\;}
\def\lap{\;\rlap{\lower 2.5pt
 \hbox{$\sim$}}\raise 1.5pt\hbox{$<$}\;}

\begin{document}

\shorttitle{The final spin from black-hole mergers}
\title{The final spin from binary black holes in quasi-circular orbits}

\author{Fabian Hofmann\altaffilmark{1}, Enrico
  Barausse\altaffilmark{2,3}, Luciano Rezzolla\altaffilmark{1,4}}

\altaffiltext{1}{Institute for Theoretical Physics, Max-von-Laue-Str. 1,
  60438 Frankfurt, Germany}

\altaffiltext{2}{Sorbonne Universit\'{e}s, UPMC Univ Paris 06, UMR 7095,
  Institut d'Astrophysique de Paris, F-75014, Paris, France}

\altaffiltext{3}{CNRS, UMR 7095, Institut d'Astrophysique de Paris,
  F-75014, Paris, France}

\altaffiltext{4}{Frankfurt Institute for Advanced Studies,
  Ruth-Moufang-Str. 1, 60438 Frankfurt, Germany}

\begin{abstract}
We revisit the problem of predicting the spin magnitude and direction of
the black hole resulting from the merger of two black holes with
arbitrary masses and spins inspiralling in quasi-circular orbits. We do
this by analyzing a catalog of 619
recent numerical-relativity simulations collected from the literature and
spanning a large variety of initial conditions. By combining information
from the post-Newtonian approximation, the extreme mass-ratio limit and
perturbative calculations, we improve our previously proposed
phenomenological formulae for the final remnant spin. In contrast with
alternative suggestions in the literature, and in analogy with our
previous expressions, the new formula is a simple algebraic function of
the initial system parameters and is not restricted to binaries with
spins aligned/anti-aligned with the orbital angular momentum, but can be
employed for fully generic binaries. The accuracy of the new expression
is significantly improved, especially for almost extremal progenitor
spins and for small mass ratios, yielding a root-mean-square error
$\sigma\approx0.002$ for aligned/anti-aligned binaries and
$\sigma\approx0.006$ for generic binaries. Our new formula is suitable
for cosmological applications and can be employed robustly in the
analysis of the gravitational waveforms from advanced interferometric
detectors.
\end{abstract}

\keywords{black hole physics, gravitational waves, gravitation}

\section{Introduction}

According to the predictions of general relativity, binary systems of
compact objects are the most efficient emitters of gravitational waves
(GWs). Indeed, Advanced LIGO has recently detected the GW signal from a
black-hole (BH) binary with masses $M_1\approx36\,M_\odot$ and
$M_2\approx29\,M_\odot$~\citep{Abbott:2016blz}, at a (luminosity)
distance of $\sim 410$ Mpc. In general, Advanced LIGO and other
terrestrial interferometers, such as Advanced Virgo and KAGRA, target BH
binaries with a variety of masses~\citep[up to a few hundred $M_\odot$,
  if they exist;][]{2014ApJ...789..120B,2016ApJ...819..108B}. More
massive BH binaries are targeted by existing pulsar-timing arrays
\citep[in the mass range $10^8-10^{10}\,M_\odot$;][]{2013CQGra..30v4010M}
and by future space-borne interferometers such as eLISA \citep[in the
  mass range $10^4-10^7\,M_\odot$;][]{elisa}.

One of the obvious difficulties of observing BH binaries with terrestrial
interferometers is that only the final part of the inspiral and the
merger/ringdown are in band. This is where the perturbative
post-Newtonian (PN) techniques valid earlier in the inspiral become
inaccurate, preventing the extraction the source's physical
parameters. Hence, to obtain the full gravitational waveforms, it is
necessary to resort to numerical-relativity (NR) simulations. In
practice, even under the reasonable assumption that BH binaries near the
merger have been circularized by earlier GW emission, the space of
parameters to be probed (the mass ratio $q$ and the spin vectors
$\boldsymbol{S}_1,~\boldsymbol{S}_2$, \ie~seven parameters) is too large
to be handled by NR simulations alone.

To ensure a sufficient coverage of the parameter space, semi-analytical
techniques allowing faster waveform production are employed, \eg~the
spin-effective-one-body (sEOB)
model~\citep{Buonanno:1998gg,2001PhRvD..64l4013D,Barausse:2009xi} or
``hybrid'' waveforms~\citep{2008PhRvD..77j4017A,2015arXiv150807253K},
which combine results from NR simulations with PN and quasinormal-mode
calculations. These techniques are faster, but require great care when
modeling the merger and the transition to the ringdown.  Indeed, although
the ringdown can be modeled via a linear superposition of quasi-normal
modes, their frequencies depend on the remnant BH's mass and
spin, which, in turn, depend on the initial binary parameters.

This relation between the binary's initial and final states is highly
non-trivial because it encodes the details of the strong-field, highly
relativistic merger, which is only accessible via NR calculations. Yet, a
number of approaches to predict analytically or semi-analytically the
remnant's final spin magnitude and direction have been proposed. These
range from modeling the GW fluxes throughout the binary's evolution
within the EOB model~\citep[\eg~by][for nonspinning
  BHs]{2007PhRvD..76d4003D} to approaches that combine information from
PN theory, the extreme mass-ratio limit (EMRL), symmetry arguments, and
fits to NR data, to provide ``formulae'' for the final
spin~\citep{Rezzolla:2007a,kesden_can_2008,Rezzolla:2007b,tichy_final_2008,Rezzolla:2007c,buonanno_estimating_2008,barausse_predicting_2009,healy_remnant_2014}. Similar
formulae have also been derived for the remnant's final
mass~\citep{tichy_final_2008,kesden_can_2008,barausse_mass_2012,healy_remnant_2014},
which differs from the binary's total mass by the energy emitted in GWs.
Again, a common problem in these attempts is the difficulty to cover with
sufficient accuracy the seven-dimensional parameter space of
quasi-circular BH binaries. Indeed, while most of these formulae formally
cover the whole parameter space, they can be rather inaccurate,
especially for BHs with almost extremal spins.

By combining results from NR and information from the EMRL and PN theory,
we here derive a new formula for the spin magnitude and direction for the
merger remnant from quasi-circular BH binaries with arbitrary masses and
spins. We calibrate our formula against a catalog of 619 
recently published NR
simulations~\citep{chu_high_2009,hannam_simulations_2010,nakano_intermediate-mass-ratio_2011,sperhake_numerical_2011,pollney_gravitational_2011,kelly_mergers_2011,buchman_simulations_2012,lovelace_high-accuracy_2012,hemberger_final_2013,hinder_error-analysis_2013,kelly_decoding_2013,pekowsky_comparing_2013,healy_remnant_2014,Lousto_2014b,lovelace_nearly_2015,scheel_improved_2015,szilagyi_approaching_2015,zlochower_modeling_2015,husa_frequency-domain_2016,SXS}\footnote{Note
  that for the simulations of \citet{zlochower_modeling_2015}, we only
  consider the horizon-extracted data, and \textit{not} the
  radiation-based ones, which may be imprecise~\citep{lousto_private}.},
and validate it by comparing its results to self-force calculations and
plunge-merger-ringdown fluxes for nonspinning binaries with small mass
ratios, as well as to a set of 248 NR simulations not included in the
calibration dataset~\citep{gatech_2016}.

Our new formula builds upon \citet{barausse_predicting_2009}, who
introduced a final-spin formula that is widely used both in the
production of semi-analytical waveforms (\eg~in sEOB and phenomenological
waveforms) and in cosmological studies of massive BH
evolution~\citep[see, \eg][]{bertispin,fanidakis,baraussespin,sikora,dubois,Sesana:2014bea}. Our novel prescription
especially improves the accuracy of the formula by
\citet{barausse_predicting_2009} for extreme mass ratios and for
near-extremal spins. This is important since near-extremal spins are
expected, at least in some cases, for supermassive BHs
\citep{bertispin,fanidakis,baraussespin,sikora,dubois,Sesana:2014bea} and
possibly also for stellar-mass BHs~\citep{2011CQGra..28k4009M}. We assume
$G=1=c$ hereafter.

\section{Modeling the final spin}

Let us first consider a BH binary with spins parallel (\ie~aligned or
anti-aligned) to the orbital angular momentum $\boldsymbol{L}$, and
denote the masses by $M_{1,2}$ (with $q\equiv M_2/M_1\leq1$) and the spin projections on the
angular-momentum direction by $S_{1,2}\equiv a_{1,2}\,M_{1,2}^2$
($a_{1,2}$ being the dimensionless spin-parameter projections). In the
EMRL $q\ll 1$, the final-spin projection on the
angular-momentum direction must be
\begin{equation}\label{emrl}  
a_{\rm fin}=a_1+\nu\left(L_{_{\rm ISCO}}(a_1)-2a_{1}E_{_{\rm ISCO}}(a_1)\right)+\mathcal{O}(\nu^2)\,,
\end{equation}   
with $\nu\equiv q/(1+q)^2$ the symmetric mass ratio, and $L_{_{\rm
    ISCO}}(a)$, $E_{_{\rm ISCO}}(a)$, respectively the specific (dimensionless) angular
momentum and energy for a test particle at the innermost stable circular
orbit (ISCO) of a Kerr BH with spin parameter $a$~\citep{Bardeen:1972fi}
\begin{eqnarray}
\label{eisco} 
&&{E}_{_{\rm ISCO}}(a)=\sqrt{1-\frac{2}{3{r}_{_{\rm ISCO}}(a)}}\,,\\ 
&&{L}_{_{\rm ISCO}}(a)=\frac{2}{3\sqrt{3}}\left[1+2\sqrt{3{r}_{_{\rm ISCO}}(a)-2}\right]\,,
\label{lisco} 
\end{eqnarray} 
\begin{eqnarray} 
&&{r}_{_{\rm ISCO}}(a)=3+Z_{2}-\frac{a}{|a|}\sqrt{(3-Z_{1})(3+Z_{1}+2Z_{2})}\,,\\ 
&&Z_{1}=1+(1-a^2)^{1/3}\left[(1+{a})^{1/3}+(1-{a})^{1/3}\right]\,,\\ 
&&Z_{2}=\sqrt{3{a}^{2}+Z_{1}^{2}}\,.
\end{eqnarray} 
The final-spin expression of \citet{Rezzolla:2007c} and
\citet{barausse_predicting_2009} reproduces Eq.~\eqref{emrl} only in the
special case $a_1=0$, when $a_{\rm
  fin}=2\sqrt{3}\nu+\mathcal{O}(\nu^2)$. Indeed, one of the drawbacks of
those early expressions is that they may yield spins $a_{\rm fin}>1$ for
small mass ratios $\nu\ll 1$, in clear disagreement with
Eq.~\eqref{emrl}, which predicts $a_{\rm fin}\leq 1$, the equality
holding for $a_1=1$.

To enforce the EMRL exactly, we consider the following ansatz for the
final-spin projection:
\begin{multline} 
a_{\rm fin}={a}_{\rm tot}+\nu[L_{_{\rm ISCO}}({a}_{\rm eff})-2{a}_{\rm tot}(E_{_{\rm
        ISCO}}({a}_{\rm eff})-1)]\\
+\sum_{i=0}^{n_{_M}}\sum_{j=0}^{n_{_J}}k_{ij}\nu^{2+i}\,{a}_{\rm  eff}^j\,, 
\label{new_ansatz} 
\end{multline} 
where $k_{ij}$ are free coefficients to be determined from the NR data,
${a}_{\rm tot}\equiv(S_1+S_2)/(M_1+M_2)^2=({a_1}+{a_2}q^2)/(1+q)^2$ is
the ``total'' spin parameter used in \citet{barausse_mass_2012}, while
$a_{\rm eff}\equiv S_{\rm eff}/(M_1+M_2)^2$ is an ``effective'' spin
parameter. In more detail, we assume $S_{\rm eff}=(1 + \xi M_2/M_1) S_1 +
(1 + \xi M_1/M_2) S_2$, which yields $a_{\rm eff}=a_{\rm tot}+\xi \nu
(a_1+a_2)$. This choice is inspired by \citet{2001PhRvD..64l4013D}, who
finds that the leading-order conservative spin-orbit dynamics depends on
the spin only through $S_{\rm eff}$ with $\xi=3/4$, while the
leading-order conservative spin-spin dynamics depends on $S_{\rm eff}$
with $\xi=1$ \citep[see also][]{racine,kesden1,kesden2}. In the
following, we will keep $\xi$ as a free parameter and determine it from
the NR simulations.\footnote{Setting $\xi=3/4$ or $\xi=1$ yields a much
  larger reduced $\chi^2$ (see below for how we compute it). For
  $n_{_M}=1,\,n_{_J}=2~(n_{_M}=3,\,n_{_J}=4)$ we obtain $\chi^2_{\rm red}
  \approx5~(1.4)$ for $\xi=3/4$, and $\chi^2_{\rm red}\approx51$ (10) for
  $\xi=1$. This strong statistical evidence that $\xi\neq3/4,\,1$ is not
  surprising, as one indeed expects the leading-order spin-orbit and
  spin-spin couplings to be ``deformed'' for highly relativistic
  binaries~\citep{Barausse:2009xi}.}

Note that Eq.~\eqref{new_ansatz} matches Eq.~\eqref{emrl} for $\nu\ll1$,
since $a_{\rm tot}=a_1(1-2\nu)+\mathcal{O}(\nu^2)$. Moreover, by singling
out ${a}_{\rm tot}$ as the first term in Eq.~\eqref{new_ansatz}, we have
isolated the ``direct'' contribution of the progenitor spins to the
remnant's spin. However, this does not mean that all leading-order
effects of the smaller BH's spin $a_2$ are already included. For
instance, the specific energy and angular momentum at the ISCO receive 
corrections of ${\cal O}(a_2\,\nu)$ \citep[see \eg][]{Barausse:2009xi},
which propagate into a term of ${\cal O}(a_2\,\nu^2)$ in the final 
spin, c.f. Eq.~\eqref{emrl}. This effect, together with other ones, 
is captured by the coefficient $k_{01}$.

The coefficients $k_{0j}$ of the $\nu^2$ terms in Eq.~(\ref{new_ansatz})
also encode the information about the self-force dynamics (both
dissipative and conservative) and the leading-order (in mass ratio)
plunge-merger-ringdown emission. More specifically, the conservative
self-force produces shifts $\nu\,\Delta E_{_{\rm ISCO}}$ and
$\nu\,\Delta L_{_{\rm ISCO}}$ in the ISCO specific energy
and angular momentum away from the geodesic values of
Eqs.~(\ref{eisco}) and (\ref{lisco}). For a nonspinning binary
($a_1=a_2=0$) with $\nu\ll 1$ 
\begin{equation} 
\Delta L_{_{\rm ISCO}}\approx-0.802\,.
\end{equation} 
This follows from evaluating Eq. (3c) of \cite{2012PhRvL.108m1103L} at
the ISCO frequency, which should include conservative self-force effects
as in Eq. (5) of the same reference. The plunge-merger-ringdown
angular-momentum flux is instead given
by~\citep{2010PhRvD..81h4056B} 
\begin{equation} 
\Delta J_{_{\rm MR}}\approx3.46\,\nu^2\,. 
\end{equation} 
Therefore, for a nonspinning binary one expects
\begin{align}
a_{\rm fin}\approx&\,\nu L_{_{\rm ISCO}}(0)-2\nu^{2}[E_{_{\rm ISCO}}(0)-1]L_{_{\rm ISCO}}(0)\nonumber\\&+\Delta L_{_{\rm
	ISCO}}\nu^2-\Delta J_{_{\rm MR}}+{\cal O}(\nu^3)\\\approx&\,2\sqrt{3}\nu-3.87\,\nu^2+{\cal O}(\nu^3)\,,\nonumber
\end{align}
hence $k_{00}\approx-3.87$. (Note that at ${\cal O}(\nu)$ and after
setting $a_1=a_2=0$, this equation reduces to Eq.~\eqref{emrl}.) However,
since the transition from inspiral to plunge does not happen exactly at
the ISCO when accounting for deviations from adiabaticity, but takes
place smoothly around the ISCO~\citep{2000PhRvD..62l4022O}, and since the
the plunge-merger-ringdown fluxes are intrinsically approximate (as it is
difficult to define unambiguously the plunge-merger-ringdown as separate
from the late inspiral), we keep $k_{00}$ as a free parameter. As it
happens, at least for $n_{_M}=1$, $n_{_J}=2$, the fitted value is
$k_{00}\approx-3.82$, which is reasonably close to the one predicted by
the considerations above\footnote{For the cases $n_{_M}=3$, $n_{_J}=3$
  and $n_{_M}=3$, $n_{_J}=4$, also considered in the following,
  $k_{00}\approx-5.9$. However, we will show that unlike $n_{_M}=1$,
  $n_{_J}=2$, those cases are probably overfitting the data.}.

In principle, we could fit all the coefficients $k_{ij}$ (as well as
$\xi)$ to the NR results. However, since simulations for equal-mass
non-spinning BH binaries have determined the final remnant's spin with
accuracy far better than for other configurations, we impose that
Eq.~(\ref{new_ansatz}) with $q=1$ and $a_1=a_2=0$ yields exactly the
final spin $a_{\rm fin}=0.68646\pm0.00004$ measured by the NR simulations
of \citet{scheel_high-accuracy_2009}. This gives the relation
\begin{equation} 
\frac{\sqrt{3}}{2}+\sum_{i=0}^{n_{_M}}\frac{k_{i0}}{4^{2+i}}=0.68646\pm0.00004\,. 
\end{equation} 
With this constraint, we fit Eq.~(\ref{new_ansatz}) to the 246
simulations for parallel-spin binaries in our calibration dataset.

\begin{table}
\begin{center}
\begin{tabular*}{\columnwidth}{cccccc}
\hline
\hline
\!\!\!$k_{01}$   & \!\!\!$k_{02}$   & \!\!\!$k_{10}$   & \!\!\!$k_{11}$  & \!\!\!$k_{12}$ &\!\!\! $\xi$ \\  
\!\!\!$-1.2019$ & \!\!\!$-1.20764$ & \!\!\!$3.79245$ & \!\!\!$1.18385$ & \!\!\!$4.90494$& \!\!\!$0.41616$ \\
\hline
\hline
& & & & & \\
\hline
\hline
\!\!\!$k_{01}$ & \!\!\!$k_{02}$ & \!\!\!$k_{03}$ & \!\!\!$k_{10}$ & \!\!\!$k_{11}$ & \!\!\!$k_{12}$ \\
\!\!\!$2.87025$ & \!\!\!$ -1.53315$ & \!\!\!$ -3.78893$ & \!\!\!$ 32.9127$ & \!\!\!$ -62.9901$ & \!\!\!$ 10.0068$ \\
\hline
\!\!\!$k_{13}$ & \!\!\!$k_{20}$ & \!\!\!$k_{21}$ & \!\!\!$k_{22}$ & \!\!\!$k_{23}$ & \!\!\!$k_{30}$ \\ 
\!\!\!$ 56.1926$ & \!\!\!$-136.832$ & \!\!\!$ 329.32$ & \!\!\!$ -13.2034$ & \!\!\!$ -252.27$ & \!\!\!$ 210.075$ \\
\hline
\!\!\!$k_{31}$ & \!\!\!$k_{32}$ & \!\!\!$k_{33}$& \!\!\!$\xi$ \\  
\!\!\!$ -545.35$ & \!\!\!$ -3.97509$ & \!\!\!$368.405$ & \!\!\!$ 0.463926$ \\
\hline
\hline
& & & & & \\
\hline
\hline
\!\!\!$k_{01} $ & \!\!\!$k_{02} $ &\!\!\! $k_{03} $ & \!\!\!$k_{04} $ & \!\!\!$k_{10} $ & \!\!\!$k_{11} $ \\ 
\!\!\!$3.39221$ &\!\!\!$4.48865$ &\!\!\!$-5.77101$ &\!\!\!$-13.0459$ &\!\!\!$35.1278$ &\!\!\! $-72.9336$ \\
\hline
\!\!\!$k_{12} $ & \!\!\!$k_{13} $ &\!\!\! $k_{14} $ &\!\!\! $k_{20} $ & \!\!\!$k_{21} $ &\!\!\! $k_{22} $ \\
\!\!\!$-86.0036$ &\!\!\!$93.7371$ &\!\!\!$200.975$ &\!\!\!$-146.822$ &\!\!\!$387.184$ &\!\!\!$447.009$ \\
\hline
 \!\!\!$k_{23} $ & \!\!\!$k_{24} $ &\!\!\!$k_{30} $ & \!\!\!$k_{31} $ & \!\!\!$k_{32} $ & \!\!\!$k_{33} $ \\
 \!\!\!$-467.383$ &\!\!\!$-884.339$ &\!\!\!$223.911$ &\!\!\!$-648.502$ &\!\!\!$-697.177$ &\!\!\!$753.738$ \\
\hline
\!\!\!$k_{34}$ &\!\!\!$\xi$ \\  
\!\!\!$1166.89$ &\!\!\!$0.474046$ \\
\hline
\hline
\end{tabular*}
\caption{The coefficients of our formula, for $n_{_M}=1,~n_{_J}=2$ (top
  block), $n_{_M}=3,~n_{_J}=3$ (middle block) and $n_{_M}=3,~n_{_J}=4$
  (bottom block).}
\label{table_coeffs}
\end{center}
\end{table}

\begin{figure*}
\begin{center}
  \includegraphics[width=\columnwidth]{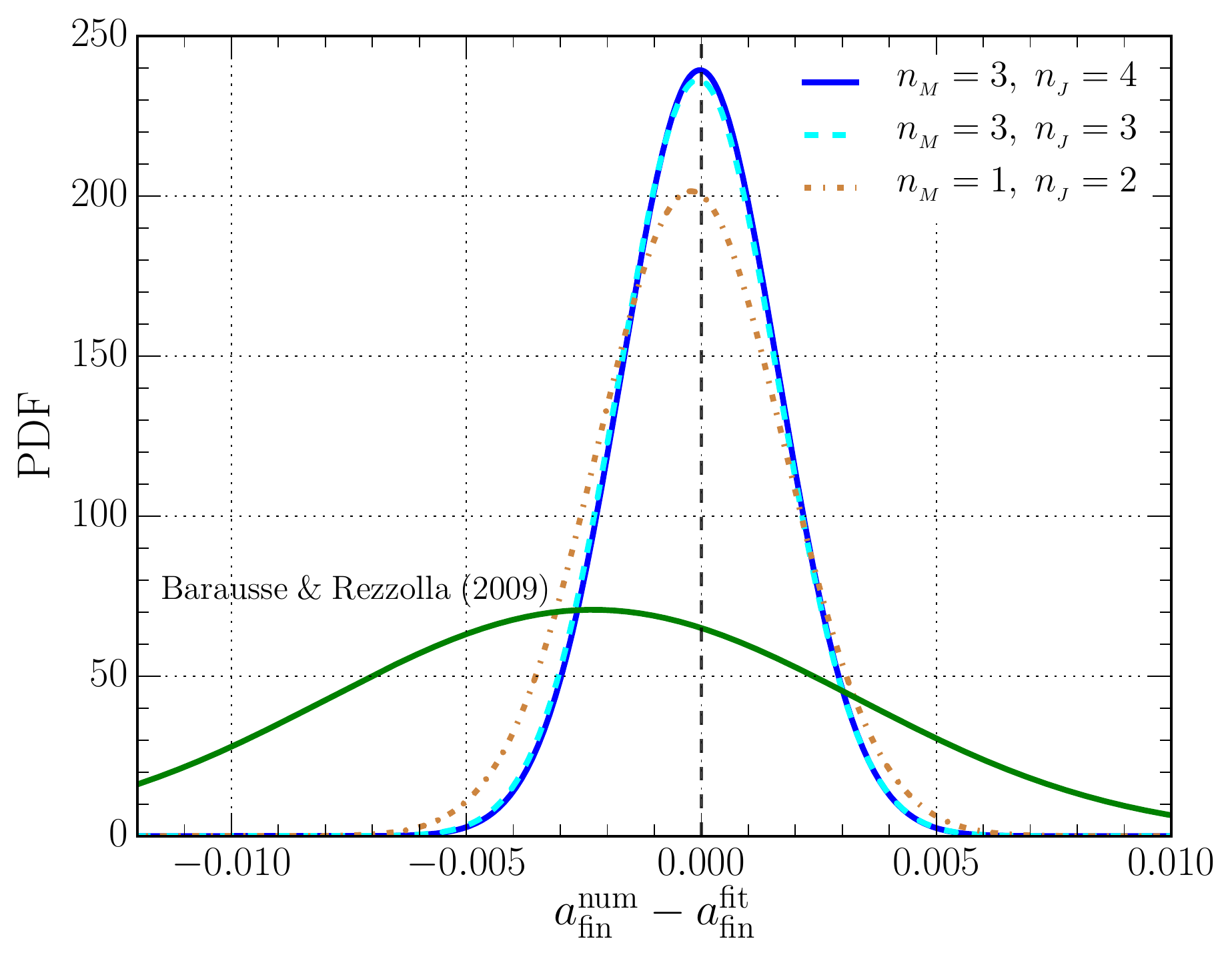}
  \includegraphics[width=\columnwidth]{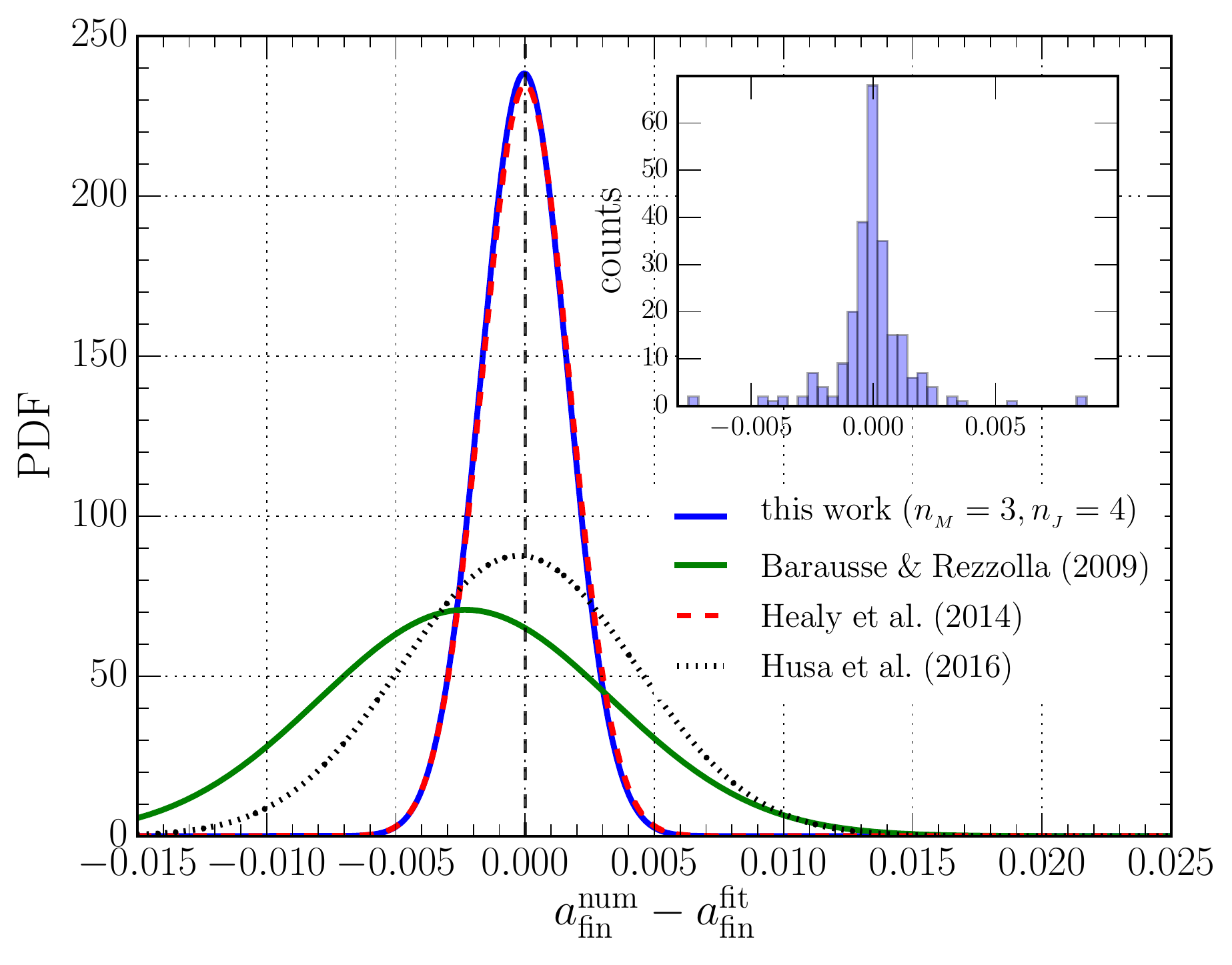}
  \caption{Left panel: probability distribution functions (PDFs),
    obtained as Gaussian fits, for the residuals of our formula with
    increasing number of coefficients
    (\ie~$n_{_M}=1,\,n_{_J}=2;~n_{_M}=3,\,n_{_J}=3;~n_{_M}=3,\,n_{_J}=4$)
    and for that of \citet{barausse_predicting_2009}. Right panel: same
    as the left panel, but for our formula with $n_{_M}=3,\,n_{_J}=4$ and
    for the formulae of \citet{barausse_predicting_2009},
    \citet{healy_remnant_2014}, and \citet{husa_frequency-domain_2016};
    the inset shows the actual distribution for our formula.}
  \label{PDFaligned}
\end{center}
\end{figure*}

However, before performing the fit, it is useful to quantify the average
error of the final spins calculated from NR simulations. This is possible
because our calibration dataset contains simulations by different groups with
the same initial data. More precisely, 71 parallel-spin simulations have
one or more ``twins'', \ie~binaries with exactly the same
initial properties, so that the \textit{mean} of the absolute differences
between twin NR simulations can be measured to be
$\delta a_{\rm fin}\approx0.002$. This estimate allows not only 
performing a fit, but also computing its reduced chi-squared $\chi^2_{\rm
  red}$, thus gauging whether we are overfitting the data, which would
correspond to $\chi^2_{\rm red}<1$.

\begin{table}
\label{table_goodness}
\begin{center}
\begin{tabular}{lcccc}
\hline
\hline
Model                       &coeffs. & $\mu$      & $\sigma$ & $\chi_{\mathrm{red}}^2$\\
\hline
$n_{_M}=1,\ n_{_J}=2$         & $\phantom{0}6$  & $-0.000215$ & $0.00198$ & $0.985$ \\
$n_{_M}=3,\ n_{_J}=3$         & $16$ & $-0.000066$ & $0.00168$ & $0.712$ \\
$n_{_M}=3,\ n_{_J}=4$         & $20$ & $-0.000029$ & $0.00166$ & $0.694$ \\
Barausse \& Rezzolla (2009) & $\phantom{0}4$  & $-0.002310$ & $0.00564$ & $9.313$ \\
Husa {et al.} (2016)        & $11$ & $-0.000240$ & $0.00453$ & $5.150$ \\
Healy {et al.} (2014)       & $19$ & $\phantom{-}0.000014$ & $0.00170$ & $0.718$ \\
\hline
\hline
\end{tabular}
\end{center}
  \caption{The mean and rms ($\mu$ and $\sigma$) of the residuals
    $a_{\mathrm{fin}}^{\mathrm{num}}-a_{\mathrm{fin}}^{\mathrm{fit}}$
    from the numerical data, as well as $\chi^2_{\rm red}$, for our
    formula and those of \citet{barausse_predicting_2009},
    \citet{husa_frequency-domain_2016}, and \citet{healy_remnant_2014}; also
    displayed is the number of coefficients in the various cases.}
\end{table}	

Since Eq. \eqref{new_ansatz} can be expanded to arbitrary order via its
last term, we have performed fits of the parallel-spin calibration
dataset with $n_{_M}=1,\ n_{_J}=2$ (6 coefficients),
$n_{_M}=3,\ n_{_J}=3$ (16 coefficients) and $n_{_M}=3,\ n_{_J}=4$ (20
coefficients). The fitted coefficients are given in Table
\ref{table_coeffs}.  Table \ref{table_goodness} reports the mean ($\mu$)
and root-mean-square (rms, $\sigma$) of the residuals from the NR data,
as well as $\chi^2_{\rm red}$, for the three aforementioned sets of
coefficients, and for the formulae of \citet{barausse_predicting_2009},
\citet{husa_frequency-domain_2016} and \citet{healy_remnant_2014} (which
use 4, 11 and 19 coefficients, respectively). Table \ref{table_goodness}
shows that our new formula converges when increasing the number of
coefficients, although the optimal choice to avoid overfitting appears to
be $n_{_M}=1,\ n_{_J}=2$. The convergence of our formula is also
displayed in the left panel of Fig.~\ref{PDFaligned}, which shows the
probability distribution functions (PDFs) of the residuals, obtained 
as Gaussian fits. The right panel shows instead the (fitted)
PDFs for our formula (for $n_{_M}=3,\ n_{_J}=4$) and for the formulae of
\citet{barausse_predicting_2009}, \citet{husa_frequency-domain_2016} and
\citet{healy_remnant_2014}; the inset shows the actual residual
distribution for our formula. Note that already with 16 coefficients our
new formula has a slightly smaller rms than \citet{healy_remnant_2014},
with the important advantage that it can be used also for generic
binaries (see below), unlike the formulae of
\citet{husa_frequency-domain_2016} and \citet{healy_remnant_2014}.
 
To generalize Eq.~(\ref{new_ansatz}) to generic spins, we write the
remnant's spin as the total spin
$\boldsymbol{S}=\boldsymbol{S}_1+\boldsymbol{S}_2$ plus an angular
momentum contribution (\ie~the angular momentum at the binary's
``effective'' ISCO), \ie~$\boldsymbol{S}_{\rm
  fin}=\boldsymbol{S}+\Delta\boldsymbol{L}$. Since the final mass is
$M_{\rm fin}=(M_1+M_2) (1-E_{\rm rad})$ \citep[with $E_{\rm rad}\lesssim
  0.1$ the mass radiated in GWs;][]{barausse_mass_2012}, the final spin
parameter is
\begin{equation}\label{afin} \boldsymbol{a}_{\rm
    fin}=\boldsymbol{a}_{\rm tot}+\boldsymbol{\ell}\nu\,,\quad
  \boldsymbol{a}_{\rm
    tot}=\frac{1}{(1+q)^2}\left(\boldsymbol{a}_1+\boldsymbol{a}_2
  q^2\right)\,, 
\end{equation} 
where we have reabsorbed the radiated energy $E_{\rm rad}$ in
$\boldsymbol{\ell}\equiv\Delta\boldsymbol{L}/[M_{1}M_{2}(1-E_{\rm
    rad})^2]+\boldsymbol{S}[2 E_{\rm rad}+3E_{\rm rad}^2+{\cal O}(E_{\rm
    rad})^3]/(M_1 M_2)$ (note that $\boldsymbol{\ell}$ remains finite in
the test-particle limit because
$|\Delta\boldsymbol{L}|=\mathcal{O}(\nu)=E_{\rm rad}$ as $\nu\to0$). By
evaluating Eq.~(\ref{afin}) for parallel spins and comparing it to
(\ref{new_ansatz}), we obtain
\begin{multline}
|\boldsymbol{\ell}|=\Bigg\vert L_{_{\rm ISCO}}({a}_{\rm eff})-2{a}_{\rm
  tot}\left(E_{_{\rm ISCO}}({a}_{\rm eff})-1\right)\\
+\sum_{i=0}^{n_{_M}}\sum_{j=0}^{n_S}k_{ij}\nu^{1+i}\,{a}_{\rm eff}^j\Bigg\vert\,, 
\label{ell} 
\end{multline} 
which can be generalized to precessing spins by following
\citet{barausse_mass_2012} \citep[see
  also][]{Rezzolla:2007c,barausse_predicting_2009} and replacing
\begin{align}
\label{atot}
a_{\rm tot}&\to  a_{\rm
  tot}(\beta,\gamma,q)\equiv\frac{|\boldsymbol{a}_1|\cos\beta+|\boldsymbol{a}_2|\cos\gamma\,q^2}{(1+q)^2}\,,\\
\label{aeff}
{a}_{\rm eff}&\to a_{\rm eff}(\beta,\gamma,q)\equiv a_{\rm
    tot}(\beta,\gamma)+\xi\nu(a_1\cos\beta+a_2\cos\gamma)\,,
\end{align}
with $\beta$ ($\gamma$) being the angle between $\boldsymbol{a}_1$
($\boldsymbol{a}_2$) and the orbital angular momentum. Clearly, with
this choice Eq.~\eqref{afin} matches Eq.~\eqref{new_ansatz} for parallel
spins (\ie~for $\beta=0,\pi$ and $\gamma=0,\pi$).

Moreover, for equal masses ($q=1$), the leading-order PN spin effects in
the conservative sector (\ie~the leading-order spin-orbit coupling) enter
the dynamics only through the combination
${\boldsymbol{\hat{L}}}\cdot\boldsymbol{S}/M^2=a_{\rm
  tot}(\beta,\gamma,1)\propto a_{\rm eff}(\beta,\gamma,1)$ \citep[see
  \eg][]{2001PhRvD..64l4013D,Barausse:2009xi}, where a ``hat'' denotes a
unit-norm vector. Therefore, at this approximation order, the binding
energy and angular momentum at the effective ISCO depend on the
spins only through $a_{\rm tot}(\beta,\gamma,1)$ (or equivalently $a_{\rm
  eff}(\beta,\gamma,1)$), as reflected in
Eqs.~\eqref{ell}--\eqref{aeff}. Similarly, in the EMRL, the leading
contributions to $|\boldsymbol{\ell}|$ come from the ISCO energy and
angular momentum of a test particle in Kerr. By construction,
$|\boldsymbol{\ell}|$ has the correct EMRL for parallel spins, but the
EMRL is also recovered approximately for generic-spin configurations, at
least at leading order in the primary-BH spin. Indeed, this happens
because the ISCO angular momentum and energy for a test particle in a
non-equatorial orbit in a Kerr spacetime are $L_{_{\rm ISCO}}(a_{\rm
  tot}(\beta,\gamma,0))$ and $E_{_{\rm ISCO}}(a_{\rm
  tot}(\beta,\gamma,0))$, at leading order in the spin~\citep[see
  discussion in][]{barausse_mass_2012}.

Putting things together, the final-spin magnitude reads
\begin{eqnarray}
&|\boldsymbol{a}_{\rm fin}|=\frac{1}{(1+q)^2}\big[|\boldsymbol{a}_1|^2+|\boldsymbol{a}_2|^2q^4+2|\boldsymbol{a}_1||\boldsymbol{a}_2|q^2\cos\alpha\notag\\ 
&+2(|\boldsymbol{a}_1|\cos{\beta}+|\boldsymbol{a}_2|q^2\cos{\gamma})|\boldsymbol{\ell}|q+|\boldsymbol{\ell}|^2q^2\big]^{1/2}\,, \label{afinnorm}
\end{eqnarray}
where $\alpha$ is the angle between the two spins. In principle, the
angles $\alpha$, $\beta$ and $\gamma$ depend on the binary
separation. However, $\beta$ and $\gamma$ enter in our formulae only
through $a_{\rm tot}(\beta,\gamma,q)$ and $a_{\rm
  eff}(\beta,\gamma,q)$. These combinations remain constant during the
adiabatic inspiral~\citep{acst-94}, if only the leading PN order in the
spins (\ie~the leading-order spin-orbit coupling) is included, and either
\textit{(i)} the masses are equal; or \textit{(ii)} only one BH is
spinning; or \textit{(iii)} the mass ratio is extreme
(\ie~$\nu\approx0$). Similarly, under the same assumptions, we can safely
assume that $\alpha$ remains constant during the adiabatic
inspiral~\citep{acst-94}, \ie~the angle between the two spins is
preserved by the leading-order spin-orbit coupling for equal masses,
while it does enter the final-spin prediction if only one BH is spinning,
or when $\nu\approx 0$ (indeed, the effect of the smaller BH's spin
vanishes at leading order in $\nu$, because
$\vert\boldsymbol{S}_2\vert={\cal O}(\nu)^2$). Outside these special
cases, $\alpha$, $\beta$ and $\gamma$ are not exactly constant. For
instance, in general $\alpha$ oscillates and the oscillations may even
become ``flip-flop''-unstable between separations $r_{\rm ud\pm}=
(\sqrt{a_1}\pm\sqrt{q\,a_2})^4(M_1+M_2)/(1-q)^{2}$ for certain
unequal-mass configurations where the primary-BH spin is aligned with the
orbital angular momentum and the spin of the secondary is anti-aligned
with it~\citep{2016arXiv160105086L,gerosa_precessional_2015}. These
configurations, however, are unlikely if the spins are isotropically
distributed, or if the spins are almost aligned with the angular momentum
of a circumbinary disk due to the Bardeen-Petterson
effect~\citep{Bardeen:1975zz}.

Therefore, we follow~\citet{barausse_mass_2012,barausse_predicting_2009,Rezzolla:2007c} and
define $\alpha,\,\beta$ and $\gamma$ at the initial binary separation
$r_{\rm in}$
\begin{align}
\cos\alpha &\equiv{\boldsymbol{\hat{a}}}_1\cdot{\boldsymbol{\hat{a}}}_2\vert_{r_{\rm
    in}}\,,\nonumber \\
\label{angles} 
\cos\beta &\equiv{\boldsymbol{\hat{L}}}\cdot
  {\boldsymbol{\hat{a}}}_1\vert_{r_{\rm in}}\,,\\
\cos\gamma &\equiv
  {\boldsymbol{\hat{L}}}\cdot{\boldsymbol{\hat{a}}}_2\vert_{r_{\rm
    in}}\,.\nonumber
\end{align} 
Indeed, \citet{barausse_predicting_2009} and \citet{2010PhRvD..81h4054K}
have verified that the final-spin predictions are robust against the
initial separation $r_{\rm in}$, \ie~in most cases the definitions
(\ref{angles}) are justified.

\begin{figure}
\begin{center}
  \includegraphics[width=\columnwidth]{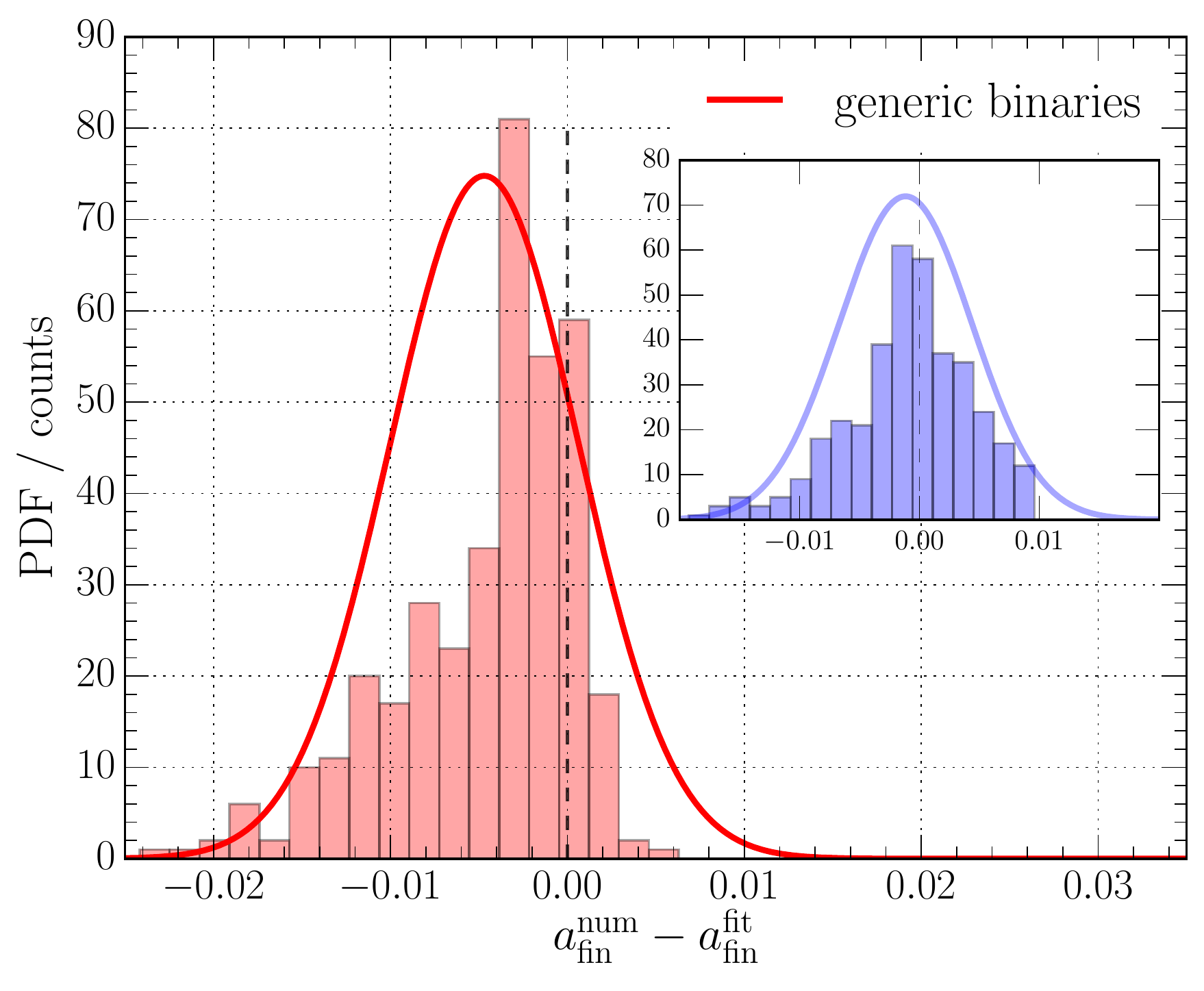}
  \caption{The residual distribution for the remnant spin magnitude, for
    binaries with generic spins; the inset shows how the modest bias of
    the distribution can be reduced by adjusting the angles
    $\beta,\,\gamma$.
  \label{PDFgeneric}}
\end{center}
\end{figure}

A comparison between our new formula with $n_{_J}=2$, $n_{_M}=1$ and the
generic-spin simulations in our calibration dataset yields the residuals
displayed in Fig.~\ref{PDFgeneric}. Also shown is the corresponding PDF
with mean $\mu\approx-0.005$ and rms $\sigma\approx0.007$. Note that in
this case we cannot reliably estimate $\chi^2_{\rm red}$, as none of the
generic-spin configurations have ``twins'' in our calibration dataset,
and the NR error is expected to be larger than in parallel-spin binaries
because of precession. Also shown by Fig.~\ref{PDFgeneric} is an
unattractive feature of our formula, namely, that the distribution of
residuals is biased toward negative values (\ie~our formula
systematically overpredicts the final spin for generic
binaries). Although this bias is small, and because it follows from
assuming that $\alpha, \beta, \gamma$ are constant, we can amend it by
replacing the angles $\alpha,\,\beta,\,\gamma$ by ``effective'' angles
$\alpha^*,\,\beta^*,\,\gamma^*$ defined as
\begin{equation}
\label{angle_map}
\Theta^*\equiv2\arctan\left[(1+\epsilon_\Theta)\tan\frac{\Theta}{2}\right]\approx\Theta+\epsilon_\Theta\sin{\Theta}\,,
\end{equation}
where $\Theta=\alpha,\,\beta,\,\gamma$, $\epsilon_\Theta$ are free
coefficients to be fixed by the data, and we impose
$\epsilon_\beta=\epsilon_\gamma$ to make our formula symmetric under
exchange of the two BHs.  Clearly, for parallel spins $\alpha^*=\alpha$,
$\beta^*=\beta$ and $\gamma^*=\gamma$. A comparison with the NR data
gives $\epsilon_\alpha\approx 0$ and
$\epsilon_\beta=\epsilon_\gamma\approx 0.024$, where we have used the
second equality of Eq.~\eqref{angle_map} (the first equality gives
similar results). The corresponding residual distribution has a smaller
bias and is shown in the inset of Fig.~\ref{PDFgeneric}, together with a
PDF with $\mu\approx-0.001$, $\sigma\approx0.006$.

As a further ``blind'' test of our formula, we consider data from the
recently published catalogue of \citet{gatech_2016} that is \textit{not}
already included in our calibration dataset (\ie~83 parallel-spin and 165
precessing-spin simulations). Already when using only $n_{_M}=1$ and
$n_{_J}=2$, the comparison yields mean and rms residuals $\mu\approx
-5\times 10^{-5}$ and $\sigma\approx1.4\times 10^{-4}$ for parallel
spins, and $\mu\approx -0.004$ ($\mu\approx -0.0005$), and
$\sigma\approx3.3\times 10^{-4}$ ($\sigma\approx3.5\times 10^{-4}$) for
precessing spins with unadjusted (adjusted) angles $\beta,\gamma$.

\begin{figure}
\begin{center}
  \includegraphics[width=\columnwidth]{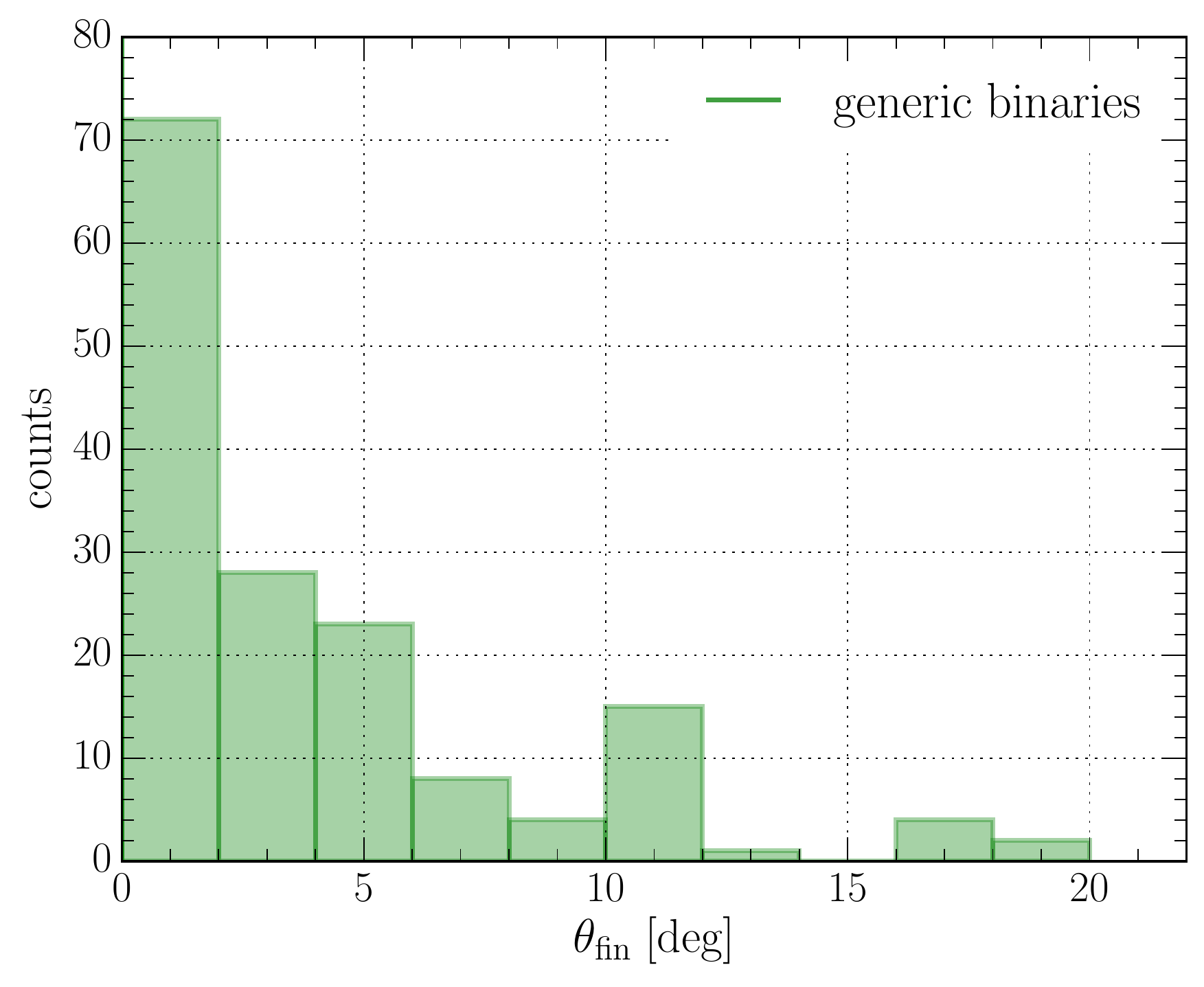}
  \caption{The distribution of the angle between the final spin and the
    initial direction of the total angular momentum.
  \label{fig_inclination}}
\end{center}
\end{figure}

Finally, for the final-spin direction we follow
\citet{barausse_predicting_2009,acst-94} and note that at leading PN
order in the spins (\ie~including the leading-order spin-orbit coupling
alone), the GW-driven evolution in the adiabatic inspiral approximately
preserves the direction of the total angular momentum
$\boldsymbol{J}\equiv\boldsymbol{L}+\boldsymbol{S}$.
\citet{barausse_predicting_2009}, and later \citet{2014PhRvD..89b1501L},
verified that $\boldsymbol{\hat{J}}$ is approximately preserved (to
within a few degrees) also in the plunge, merger and ringdown. The only
exception to this ``simple-precession'' picture are binaries with spins
almost anti-aligned with the orbital angular momentum at large
separations~\citep{acst-94,2010PhRvD..81h4054K}. Indeed, when the GW
emission sheds enough angular momentum that $\boldsymbol{L}\approx
-\boldsymbol{S}$, these binaries undergo ``transitional
precession''~\citep{acst-94}, whereby the direction of
${\boldsymbol{\hat{J}}}$ changes significantly on short timescales. Note
that among the configurations that give rise to ``simple precession'' are
also the ``flip-flop'' binaries of~\citet{2016arXiv160105086L} and
\citet{gerosa_precessional_2015}. Since transitional-precession
configurations comprise a small portion of the parameter
space~\citep{2010PhRvD..81h4054K}, we follow
\citet{barausse_predicting_2009} and assume that the final-spin direction
is simply given by ${\boldsymbol{\hat{J}}}(r_{\rm in})$, \ie~the
final-spin angle $\theta_{\rm fin}$ relative to the initial angular
momentum is simply
\begin{equation}
\label{eq:dir}
\cos\theta_{\rm fin}={\boldsymbol{\hat{J}}}(r_{\rm in})\cdot{\boldsymbol{\hat{L}}}(r_{\rm in})\,. 
\end{equation} 
Indeed, the 157 simulations~\citep{zlochower_modeling_2015,Lousto_2014b}
in our dataset that report the final-spin direction confirm that the
final spin is almost aligned with ${\boldsymbol{\hat{J}}}(r_{\rm in})$,
to within $\sim 18^\circ$ in the worst case, and to within $4^\circ$
($6^\circ$) in $64\%$ ($78\%$) of the cases. The distribution of the
angle between the final spin and ${\boldsymbol{\hat{J}}}(r_{\rm in})$ is
shown in Fig.~\ref{fig_inclination}; it is unclear whether the small
counts for $\theta_{\rm fin} \gtrsim 10^\circ$ are due to imprecisions in the
formula or in the numerical simulations.

Finally, we note that unlike other formulae for the final-spin direction
\citep{buonanno_estimating_2008,tichy_final_2008,Rezzolla:2007c},
Eq.~(\ref{eq:dir}) is valid also when $r_{\rm in}\gg M_1+M_2$. (This is also the
case for our formula for the final-spin magnitude.) This is
particularly important to predict the final spin in massive BH
mergers. Indeed, cosmological simulations (both numerical and
semi-analytical ones) cannot follow the evolution of massive BH binaries
below the separation $r_{_{\rm GW}}$ at which the GW dynamics starts
driving the orbital evolution. For a binary with
$M_1+M_2\sim10^8\,M_\odot$ in a gas-poor environment, $r_{_{\rm
    GW}}\sim10^{-2}\,{\rm pc}\sim2\times10^{3}\,(M_1+M_2)$, a separation
at which other prescriptions for the final-spin direction become
significantly inaccurate \citep[see discussion
  in][]{barausse_predicting_2009,2010JPhCS.228a2050B}.\footnote{In a gas-rich environment,
  the separation $r_{_{\rm GW}}$ below which GWs dominate 
  the binary evolution and our formulae can be applied
  is smaller \citep{Armitage_2012}, while for ``flip-flop'' binaries our
  formula for the final-spin magnitude might be applicable only below $r_{\rm ud\pm}$.}

\section{conclusion}
By combining information from the test-particle limit,
perturbative/self-force calculations, the PN dynamics, and an extensive
set of NR simulations collected from the literature, we have constructed
a novel formula for the final spin from the merger of quasi-circular BH
binaries with arbitrary mass ratios and spins. When applied to
parallel-spin configurations, our novel formula performs better than
other expressions in the literature, and we have also tested its validity
for precessing-spin binaries, which other formulae are not able to model
accurately.  Also, unlike models such as that of
\citet{healy_remnant_2014}, our formula is purely algebraic. Finally, we
have used our collected NR dataset to confirm that the final-spin
direction is almost parallel to the initial total angular-momentum
direction, as first suggested by \citet{barausse_predicting_2009}.

\acknowledgements 
We thank Nathan Johnson-McDaniel for useful comments and Davide Gerosa
for clarifications on flip-flop binaries. We acknowledge
support from the European Union's Seventh Framework Programme
(FP7/PEOPLE-2011-CIG) through the Marie Curie Career Integration Grant
GALFORMBHS PCIG11-GA-2012-321608, from the H2020-MSCA-RISE-2015 Grant
No. StronGrHEP-690904, and from the ERC Synergy Grant ``BlackHoleCam -
Imaging the Event Horizon of Black Holes'' (Grant 610058).

\newpage

\bibliographystyle{apj}
%\bibliography{ref.bib}

\end{document}